\documentstyle[12pt,english]{article}
\newenvironment{eq}
{\[\begin{array}}{\end{array}\]{}}

\let\rvec=\vec        

   \def\({\Bigl(}
\def\){\Bigr)}    \def\|{\Big|}
\def\then{~\Rightarrow~}   \def\o{\circ}
\def\m{\bullet}    \def\x{\times}
   \def\ox{\otimes}
 
\def\pl{\oplus}

\def\SUM{\displaystyle \sum}

\def\mid{\big\bracevert}

\def\sub{\subseteq}
\def\subnoteq{\subset}

\def\supnoteq{\supset}
\def\and{\wedge}

\def\AND{\displaystyle\bigwedge}

\def\rin{{\,\in\kern-.42em\in}}

\def\rep{\,{\rm rep}\,}

\def\det{\,{\rm det }\,}

\def\centr{\,{\rm centr}\,}

\def\X{\hbox{\Large$\times$}}

\def\A{{\,{\rm A\kern-.55emA}}}
\def\B{{\,{\rm I\kern-.2emB}}}
\def\C{{\,{\rm I\kern-.55emC}}}
\def\E{{\,{\rm I\kern-.2emE}}}
\def\G{{\,{\rm I\kern-.55emG}}}
\def\H{{\,{\rm I\kern-.2emH}}}
\def\I{{\,{\rm I\kern-.2emI}}}
\def\K{{\,{\rm I\kern-.2emK}}}
\def\L{{\,{\rm I\kern-.2emL}}}
\def\M{{\,{\rm I\kern-.16emM}}}
\def\N{{\,{\rm I\kern-.16emN}}}
\def\Q{{\,{\rm I\kern-.5emQ}}}
\def\R{{\,{\rm I\kern-.2emR}}}
\def\S{{\,{\rm I\kern-.42emS}}}
\def\T{{\,{\rm I\kern-.37emT}}}
\def\Z{{\,{\rm Z\kern-.35emZ}}}

\def\al{\alpha}  \def\be{\beta} \def\ga{\gamma}
\def\de{\delta}    
    \def\io{\iota}
   \def\la{\lambda}   
    
\def\phi{\varphi}

\def\mod#1{\underline{\bf mod}_{#1}}

\def\vec#1{\underline{\bf vec}_{#1}}

\def\GL{{\bf GL}}  
\def\SL{{\bf SL}}
\def\U{{\bf U}} 
\def \UL{{\bf UL}} 
   
\def\SU{{\bf SU}} 
\def\SO{{\bf SO}}

\def\norm#1{\parallel #1\parallel}

\def\angle#1{\langle#1\rangle}

\def\ro#1{{\rm #1}}
\def\bl#1{{\bf {#1}}}

\def\sprod#1#2{\langle#1|#2\rangle}

\def\map{\longrightarrow}
\def\inmap{\hookrightarrow}

\def\mape{\longmapsto}

\def\mod{{\rm mod}\,}
 
\begin{document}

\begin{titlepage} 

\hfill MPI-PhT/98-14
\vskip3cm
\centerline{\bf THE CENTRAL CORRELATIONS OF}
\centerline{\bf HYPERCHARGE, ISOSPIN, COLOUR AND CHIRALITY}
\centerline{\bf IN THE STANDARD MODEL}
\vskip2cm
\centerline{
Heinrich Saller\footnote{\scriptsize 
e-mail adress: saller@mppmu.mpg.de} 
}
\centerline{Max-Planck-Institut f\"ur Physik and Astrophysik}
\centerline{Werner-Heisenberg-Institut f\"ur Physik}
\centerline{M\"unchen}
\vskip25mm

\centerline{\bf Abstract}
The correlation of the
fractionally represented hypercharge group with the 
isospin and colour
group  in the standard model determines as faithfully represented internal group
the quotient group  ${\U(1)\x\SU(2)\x\SU(3)\over\Z_2\x\Z_3}$.
The discrete cyclic central abelian-nonabelian internal correlation 
involved is considered with respect
to its consequences for 
the representations by the  standard model fields, the electroweak mixing 
angle and the symmetry breakdown. There exists a further 
discrete  $\Z_2$-correlation between chirality and Lorentz properties and  
also a continuous $\U(1)$-external-internal 
one between hyperisospin and chirality.

\end{titlepage}

\newpage

\tableofcontents

\newpage

\advance\topmargin by -1.6cm

\section{Nonabelian Synchronization of Hypercharge}

The  standard model of the electroweak  and strong 
interactions in its minimal form \cite{WEIN} shows 
a `global' relation between the internal abelian hypercharge properties and 
the
nonabelian isospin-colour ones as described  in
 \cite{HUCK,S921,AGR}. In the following, the results of these papers
are used and, sometimes for convenience, reformulated.   

The fundamental standard model fields  transform with
irreducible representations
$[2J_L|2J_R]$ of the external Lorentz group
$\SL(\C^2 )$ and irreducible representations 
$[6y]$, $[2T]$ and $[2C_1,2C_2]$ of hypercharge $\U(1)$
(rational hypercharge number $y$), isospin $\SU(2)$ 
(integer or halfinteger isospin $T$) and colour
$\SU(3)$ as given in the following table
\begin{eq}{l}
\begin{array}{|c||c|c|c|c|c|}\hline
\hbox{\bf field}&\hbox{\bf symbol}&\SL(\C^2 )&\U(1)&\SU(2)&\SU(3)\cr
              &\bl\Psi &[2J_L|2J_R]& y &[2T]& [2C_1,2C_2]\cr\hline\hline
\hbox{left lepton}&\bl l&[1|0]&-{1\over2}&[1]&[0,0]\cr\hline
\hbox{right lepton}&\bl e&[0|1]&-1&[0]&[0,0]\cr\hline
\hbox{left quark}&\bl q&[1|0]&{1\over6}&[1]&[1,0]\cr\hline
\hbox{right quarks}&\bl u,\bl d&[0|1]&{2\over3},-{1\over 3}&[0]&[1,0]\cr\hline
\hbox{Higgs}&\bl H&[0|0]&-{1\over2}&[1]&[0,0]\cr\hline
\hbox{hypercharge gauge}&\bl A&[1|1]&0&[0]&[0,0]\cr\hline
\hbox{isospin gauge}&\bl B&[1|1]&0&[2]&[0,0]\cr\hline
\hbox{colour gauge}&\bl G&[1|1]&0&[0]&[1,1]\cr\hline
\end{array}\end{eq}With respect to the Lorentz group, $[0|0]$
designates scalar fields, $[1|0]$ and $[0|1]$
are left and right handed spinor fields resp., $[1|1]$  vector fields.
The external and internal multiplicity
(singlet, doublet, triplet, quartet, octet, etc.)
 of the Lorentz-group,
isospin and colour  representations can be computed from the
natural numbers $2J_{L,R},2T,2C_{1,2}\in\N=\{0,1,\dots,\}$
\begin{eq}{rl}
N_{\rm ext}(\bl\Psi)&=(2J_L+1)(2J_R+1)\cr
N_{\rm int}(\bl\Psi)&= N_{\rm iso}(\bl\Psi)N_{\rm col}(\bl\Psi),~~\left\{
\begin{array}{rl}
N_{\rm iso}(\bl\Psi)&=2T+1\cr
N_{\rm col}(\bl\Psi)&={(2C_1+1)(2C_2+1)(2C_1+2C_2+2)\over2}\end{array}\right.
\end{eq}Fields and antifields have reflected quantum numbers
\begin{eq}{l}
\begin{array}{|c||c|c|c|c|}\hline
\bl \Psi& [2J_L|2J_R]&y&[2T]&[2C_1,2C_2]\cr\hline
\bl \Psi^*& [2J_R|2J_L]&-y&[2T]&[2C_2,2C_1]\cr\hline
\end{array}
\end{eq}

The correlation of the
external with the internal properties will be discussed in section 7
after the discussion of the internal ones.

If a nontrivial 
hypercharge $y$ in the normalization above for
the fundamental fields with  different antifields $\bl\Psi\ne\bl\Psi^*$
is written as a rational $y={Z(y)\over N(y)}$ with integer nominator
$Z(y)\in\Z$  and natural  denominator $N(y)\in\N$, 
where $Z(y)$ and $N(y)$ have no common nontrivial divisor, 
the internal multiplicity coincides with the hypercharge fractionality
$N(y)=N_{\rm int}$, e.g. $6=N_{\rm int}(\bl q)$.

Using $6y$ in the normalization above
as elements of the cyclic groups 
$\Z_n=\Z/n\Z$ for $n=2$ (isospin) and $n=3$ (colour)   
\begin{eq}{l}
\begin{array}{|c|c|c|c|}\hline
&6y&6y\mod2&6y\mod3\cr\hline\hline
\bl l&-3&1&0\cr\hline
\bl e&-6&0&0\cr\hline
\bl q&1&1&1\cr\hline
\bl u,\bl d&4,-2&0&-1\cr\hline
\bl H&-3&1&0\cr\hline
\bl A,\bl B,\bl G&0&0&0\cr\hline
\end{array}\end{eq}the correlation between hypercharge and internal nonabelian
properties reads
\begin{eq}{rl} 
\hbox{for isospin }\SU(2):&6y\mod 2=2 T\mod 2\cr
\hbox{for colour }\SU(3):&6y\mod 3= 2(C_1-C_2)\mod 3
\end{eq}This can be formalized as follows: 
The fractional hypercharge numbers  $y$ reflect the representations of the 
centrum $\I(2)\x\I(3)$ (`David star group'
as direct product of two cyclic groups with prime order 
- the real sign group $\I(2)$ and the 
complex `Mercedes star group' $\I(3)$)
 of the isospin-colour group $\SU(2)\x\SU(3)$ 
 \begin{eq}{l}
\centr\SU(n)\cong\I(n)=\{\exp 2\pi i{r\over n}\mid r=0,1,\dots,n-1\}
\cong\Z_n,~~n\ge1\cr
\centr [\SU(2)\x\SU(3)]\cong\I(2)\x\I(3)\cong\I(6)
\end{eq}The endomorphisms of the cyclic group $\I(n)$ (for $n$ prime
simple groups, even fields)  are
determined by the mapping of the cyclic element $\exp {2\pi i\over n}$
\begin{eq}{l}
\io^r:\I(n)\map \I(n) ,~~\exp{2\pi i\over n}\mape\exp{2\pi i \over n}r,
~r=0,\dots,n-1
\end{eq}nontrivial for $r\ne0$ and
faithful (injective), i.e. $\I(n)$-automorphisms, if
$r$ and $n$ are relatively prime, i.e. 
naturals $r,n\ne0$ with only 1 as common divisor. 
All $\I(n)$-subgroups $\I(d(n))$ arise with  
as $n$-divisor $d(n)$, for $n$ prime
only $\I(1)=\{1\}$ and $\I(n)$. 
In the endomorphisms $\io^r$ they come both as images
$\io^r[\I(n)]$ and  as kernels $
\I(r(n))\cong \I(n)/\io^r[\I(n)]$. The irreducible isospin and
colour representations, denoted by $\rep\SU(2)$ and $\rep\SU(3)$ resp.,
 represent the centrum as follows
\begin{eq}{lll}
[2T]\in\rep\SU(2):
&\exp\pi i\mape\exp\pi i\cdot 2 T\mod 2\cr
[2C_1,2C_2]\in\rep\SU(3):
&\exp{2\pi i\over3}\mape 
\exp {2\pi i\over3}\cdot2(C_1-C_2)\mod 3\cr 
\end{eq}i.e. the nontrivial irreducible ones  are faithful as follows
\begin{eq}{l}
[2T]\in\rep\SU(2)\hbox{ for}
\left\{\begin{array}{ll}
\SU(2)/\I(2)\cong\SO(3)&\iff 2T\mod 2=0\cr
\SU(2)&\iff 2T\mod2=1
\end{array}\right.\cr 
[2C_1,2C_2]\in\rep\SU(3)\hbox{ for}
\left\{\begin{array}{ll}
\SU(3)/\I(3) &\iff  2(C_1-C_2)\mod 3=0\cr
\SU(3) &\iff 2(C_1-C_2)\mod 3=\pm1
\end{array}\right.
\end{eq}

Therewith, the `synchronization' 
of the nonabelian isospin-colour center $\I(2)\x\I(3)$
with the hypercharge property shows that the group, faithfully 
represented in the standard model, is the quotient group 
\begin{eq}{l}
\U(2\x 3)\cong{\U(1)\x\SU(2)\x\SU(3)\over \I(2)\x\I(3)}
\end{eq}The universal covering group for the Lie algebra of $\U(2\x 3)$ is
$\R\x\SU(2)\x\SU(3)$ with the maximal compact subgroup $\U(1)\x\SU(2)\x\SU(3)$. 
Hypercharge is centrally connected with isospin and colour.
Similar to the  full unitary
group $\U(n)$, $n\ge2$, which is a composition product $\o$ 
of two normal subgroups,
but not a  direct product $\x$
\begin{eq}{l}\begin{array}{l}
\U(n)=\U(\bl 1_n)\o\SU(n)\cr
\U(\bl 1_n)\cap\SU(n)\cong\I(n)\end{array}\then\left\{
\begin{array}{rl}
\U(n)/\U(\bl 1_n)&\cong\SU(n)/\I(n)\cr
\U(n)/\SU(n)&\cong\U(1)/\I(n)\cong\U(1)\end{array}\right.
\end{eq}the group $\U(2\x3)$, defined in $\U(6)$ with 
Pauli and Gell-Mann matrices 
$\{\tau^a\}_{a=1}^3$ and
$\{\la^j\}_{j=1}^8$ resp.
\begin{eq}{l}
\exp i[{1\over6}\al\bl 1_2\ox\bl 1_3+{1\over2}\be_a\tau^a\ox\bl1_3+
{1\over2}\ga_j\bl 1_2\ox\la^j]\in\U(2\x3)
\end{eq}is a product $\o$ of two normal subgroups, where
the nonabelian one is a direct product of two normal subgroups
\begin{eq}{rl}
\U(2\x 3)&=\U(\bl 1_6)\o[\SU(2)\ox\bl1_3\x\bl1_2\ox \SU(3)]\cr
& \U(\bl 1_6)\cap\SU(2)\ox\bl 1_3\cong\I(2)\cr
&\U(\bl 1_6)\cap\bl 1_2 \ox\SU(3)\cong\I(3)\cr
&\U(\bl 1_6)\cap[\SU(2)\ox\bl1_3\x\bl1_2\ox \SU(3)]\cong\I(2)\x\I(3)\cr
&\then\left\{
\begin{array}{rl}
\U(2\x3)/\U(\bl 1_6)&\cong\SU(2)/\I(2)\x\SU(3)/\I(3)\cr
{\U(2\x3)\over\SU(2)\ox\bl1_3\x\bl1_2\ox\SU(3)}&\cong\U(1)\end{array}\right.
\end{eq}Having nontrivial
invariant subgroups, the groups $\U(n)$ or $\U(2\x3)$ are not simple, but they
have no nontrivial direct factor. 

The common cyclic subgroup of
hypercharge and isospin-colour 
defines a `unification' quite different from the grand unification schemes,
which embed  $\U(1)\x\SU(2)\x\SU(3)$ or $\U(2\x3)$
into a larger Lie group, e.g. $\SU(5)$ or $\SO(10)$. The minimal real 
12-dimensional Lie algebra 
of the standard model remains unchanged by the discrete 
$\I(6)$-correlation. The global properties of the group - not the Lie algebra
local ones are relevant.
As known from the bound states of the nonrelativistic hydrogen atom with the
angular momentum and perihel conservation indicating a 
covering symmetry \cite{FOCK,TH3} 
$\SU(2)\x\SU(2)$ and 
the orthogonality of angular momentum and perihel vector imposing the
cyclic centrum correlation ${\SU(2)\x\SU(2)\over\I(2)}\cong\SO(4)$, also 
such discrete
`unifications' have strong consequences,  as seen in
the `square degenerated'  representation spectrum
of the hydrogen atom bound states.

\section{Fundamental and Cyclic Representations}

Before the interpretation of the fundamental standard model fields as
$\U(2\x3)$-representations (section 3) it is useful to give the definition
of fundamental and cyclic representations.

Any simple Lie algebra $L_r$ of rank $r$, e.g. the rank $(n-1)$ Lie algebra
$\log\SU(n)$ of $\SU(n)$, has $r$ fundamental representations.
Each of the  $r$ vertices in the $L_r$-Dynkin diagram is uniquely 
associated to
a root of the Lie algebra and 
then also to that fundamental representation whose
heighest weight is not orthogonal to this root.
The heighest weight of any irreducible $L_r$-representation
is a unique positive integer linear combination of the 
highest weights of the fundamental representations \cite{FULHAR,HEL,RAIF}.

E.g. the $(n-1)$ fundamental representations
for the Lie algebra $\log\SU(n)$ and, with the same notation,
for the group $\SU(n)$
 \begin{eq}{l}
\underbrace{[0,\dots,0,1,0,\dots,0]}_{r{\rm-th~ place}},~~r=1,\dots,n-1 
\end{eq}acting on complex ${n\choose r}$-dimensional vector spaces,
combine
all irreducible representations
$[2J_1,\dots,2J_{n-1}]$ using natural numbers $2J_r$. The $(n-1)$ fundamental
representations reflect the $(n-1)$ nontrivial unit 
roots in $\I(n)$.
As representation of the covering group $\SU(n)$ they are faithful
for $\SU(n)/\I(r(n))$ with $\I(r(n))$
defined in section 1.

In the additive monoid for the equivalence classes of irreducible
representations for the group $\SU(n)$
\begin{eq}{l}
\rep\SU(n)=\{[2J_1,\dots,2J_{n-1}]\mid 2J_r\in\N\}\cong\N^{n-1}
\end{eq}there are submonoids for the
irreducible representations of the quotient groups
$\SU(n)/\I(d(n))$. Any
representation defines its
$n$-ality (triality \cite{BIED} for $\SU(3)$) by
\begin{eq}{l}
r=\left({\SUM_{s=1}^{n-1}}s 2J_s\right)\mod n \in\Z_n
\end{eq}This
leads immediately to the submonoids $\rep \SU(n)/\I(r(n))\sub\rep \SU(n)$
for the  representations of
the locally isomorphic quotient groups.

The submonoid with trivial $n$-ality $r=0$ characterizes the irreducible
representations of the adjoint group $\SU(n)/\I(n)$ where the isospin and
colour gauge fields are members of for $n=2$ and $n=3$ resp.

To avoid misunderstandings with respect to the  
 embeddings of the 
 quotient group representation monoids $\rep \SU(n)/\I(r(n))\sub\rep \SU(n)$, exemplified
with  the rotation and the spin group 
$\rep \SO(3)\subnoteq \rep\SU(2)$: Any $\SO(3)$-re\-pre\-sen\-ta\-tion can be
considered as an
$\SU(2)$-representation, trivial for $\I(2)$. However, the halfinteger
$\SU(2)$-representations, e.g. $[1]$, sometimes called
`2-valued  representations', are not admitted as $\SO(3)$-representations.  
By definition,  a representation as a mapping
has to be unique.

The positive factor $2J_r$ in the combinations of the 
$r$ fundamental representations of a simple Lie algebra $L_r$
is connected with the totally symmetric power of the $r$-th fundamental  
representations.  Under the fundamental representations there are 
distinguished  cyclic fundamental representations, 
maximally three independent ones.
They generate by totally antisymmetric powers
all fundamental representations.
E.g. the simple Lie algebra
$\log \SU(n)$ has one independent cyclic 
fundamental representation given by a representation 
$[0,\dots,0,1,0,\dots,0]$ where $r$ and $n$ are relatively prime,
e.g. by the defining complex $n$-dimensional one
$[1,0,\dots,0]$. The other fundamental
representations are isomorphic to the totally antisymmetrized powers of the
defining one
\begin{eq}{rl}
{\AND^r}[1,0,\dots,0]&=
\underbrace{[0,\dots,0,1,0,\dots,0]}_{r{\rm-th~ place}}
,~~r=1,\dots,n-1\cr 
{\AND^n}[1,0,\dots,0]&=[0,\dots,0]
\end{eq}The relation to the centrum $\I(n)$ is obvious, 
$\exp{2\pi i\over n}r=\(\exp {2\pi i\over n}\)^r$.

{\scriptsize
In general, for any simple Lie algebra $L_r$,
one can take  
as cyclic fundamental representations  a subset of the representations
at the 
maximally three `loose ends' of its Dynkin diagram. 
`Entering' the diagram step by step from one `loose end',
one finds  all the other fundamental representations
in the totally antisymmetric 1st, 2nd etc. power
of the `loose end' representation.
If one encounters a double line -
for the exceptional ${\bf F}_4$, the orthogonal $\log\SO(2n+1)$ and
symplectic $\log {\bf Sp}(2n)$ - or a branching vertex - for $\log\SO(2n)$ and
the exceptional ${\bf E}_{6,7,8}$ - one might have to stop the journey. 
For $\bf G_2$ the 7-dimensional representation is cyclic. 
Since the Dynkin diagrams for
$\log\SU(n)$ have single lines only and no branching vertex, the diagram
can be `gobbled up' from any of the two  loose ends.}  

The $(n-1)$ fundamental representations for $\SU(n)$
are cyclic representations for the quotient groups $\SU(n)/\I(r(n))$, e.g.
\begin{eq}{l}
\rep\SU(4):\left\{
\begin{array}{rl}
[1,0,0]\hbox{ or }[0,0,1]&\hbox{for }\SU(4)\cr
[0,1,0]&\hbox{for }\SU(4)/\I(2)\cr
\end{array}\right.\cr
\rep\SU(6):\left\{\begin{array}{rl}
[1,0,0,0,0]\hbox{ or }[0,0.0,0,1]&\hbox{for }\SU(6)\cr
[0,1,0,0,0]\hbox{ or }[0,0.0,1,0]&\hbox{for }\SU(6)/\I(2)\cr
[0,0,1,0,0]&\hbox{for }\SU(6)/\I(3)\cr
\end{array}\right.
\end{eq}For the adjoint group $\SU(n)/\I(n)$, the
 real $(n^2-1)$-dimensional representation is cyclic
\begin{eq}{rll}
\hbox{cyclic }[2]&\in\rep\SU(2)/\I(2)&\subnoteq\rep\SU(2)\cr
\hbox{cyclic }[1,0,\dots,0,1]&\in\rep\SU(n)/\I(n)&\subnoteq\rep\SU(n),~n\ge3\cr
\end{eq}

\section{Fundamental Standard Fermion Fields as\\
Fundamental $\U(2\x3)$-Representations}

The equivalence classes of the irreducible 
representations of the abelian phase group $\U(1)$ 
are characterized by the integer winding numbers
\begin{eq}{l}
\rep\U(1)=\{[z]\mid z\in\Z\}
\end{eq}There are two fundamental representations $[\pm 1]$,
the faithful  defining ones, for the rank 1 group $\U(1)$ which 
combine all irreducible representations by
positive integer multiples $n[\pm 1]=[\pm n]$. 
They are realized for hypercharge $\U(1)$ by the lepton
fields $\bl e,\bl e^*$ 
\begin{eq}{c}
\begin{array}{|c|c|}\hline
&\U(1)\cr\hline\hline
\bl e^*&[1]\cr\hline 
\end{array}~\hbox{ and }~
\begin{array}{|c|c|}\hline
&\U(1)\cr\hline\hline
\bl e&[-1]\cr\hline 
\end{array}\cr
\cr
\hbox{fundamental }y=\pm 1
\end{eq}

Correspondingly, 
$2n$ fundamental representations   will be defined for 
the full unitary group $\U(n)$
\begin{eq}{rl}
[{r\over n}||\underbrace{0,\dots,0,1,0,\dots,0}_{r{\rm-th~ place}}]
,~~r=1,\dots,n-1,&\hbox{and }[1||0,\dots,0]\cr 
[-{r\over n}||\underbrace{0,\dots,0,1,0,\dots,0}_{(n-r){\rm-th~ place}}]
,~~r=1,\dots,n-1,&\hbox{and }[-1||0,\dots,0]\cr 
\end{eq}They are 
the  antisymmetric powers of two cyclic representations
\begin{eq}{rl}
{\AND^r}[{1\over n}||1,0,\dots,0]
&=[{r\over n}||0,\dots,0,1,0,\dots,0],~r=1,\dots,n-1\cr 
{\AND^n}[{1\over n}||1,0,\dots,0]&=[1||0,\dots,0]\cr
{\AND^r}[-{1\over n}||0,0,\dots,1]&=
[-{r\over n}||0,\dots,0,1,0,\dots,0],~r=1,\dots,n-1\cr
{\AND^n}[-{1\over n}||1,0,\dots,0]&=[-1||0,\dots,0]\cr
\end{eq}and act on complex ${n\choose r}$-dimensional vector spaces.
All irreducible $\U(n)$-representations are given by
\begin{eq}{c} 
\rep\U(n)=\{
[y||2J_1,\dots,2J_{n-1}]
\mid y=z+
{1\over n}{\SUM_{r=1}^{n-1}}r 2J_r,~z\in\Z,~2J_r\in\N\}\cr
[y||2J_1,\dots,2J_{n-1}]^*=
[-y||2J_{n-1},\dots,2J_1]\cr
\end{eq}with a rational winding number 
$y\in{1\over n}\Z$ for $\U(\bl 1_n)$ in $\U(n)$.

The $\U(n)$-representation monoid can be embedded 
with a hypercharge renormalization (multiplication with $n$) 
as a  submonoid of the representation monoid for the
direct product group,  as a true submonoid  for the nonabelian case
\begin{eq}{l}
\rep\U(n)\inmap\rep[\U(1)\x\SU(n)]=\rep\U(1)\x\rep\SU(n)
\end{eq}

The two cyclic fundamental $\U(n)$-representations combine the cyclic one 
for representations of the adjoint 
group $\U(n)/\centr\U(n)\cong\SU(n)/\I(n)$
\begin{eq}{rl}
\hbox{for $\SU(2)$}:&[{1\over2}|1]
\ox [-{1\over2}|1]\cong[0|2]\cr
n\ge3:&[{1\over n}||1,0,\dots,0]\ox[-{1\over n}||0,\dots,0,1]\cong
[0||1,0,\dots,0,1]
\end{eq}

E.g. for  $\U(2)$, there are 
four fundamental representations 
\begin{eq}{l}
\begin{array}{l}
\exp i[{1\over2}\al\bl1_2+{1\over2}\be_a\tau^a]\cr
\exp i\al\end{array}~~\hbox{ and }~~
\begin{array}{l}
\exp i[-{1\over2}\al\bl1_2-{1\over2}\be_a\tau^a]\cr
\exp -i\al\end{array}\cr
\end{eq}which are realized for  hyperisospin by the $2\cdot 2$ 
fundamental lepton fields of the standard model
with isospin $T={1\over2},0$
 \begin{eq}{c}
\rep\U(2)=\{[y||2T]\mid y=z+T,~ z\in\Z,~2T\in\N\}\cr\cr
\begin{array}{|c|c|}\hline
&\U(2)\cr\hline\hline
\bl l^*&[{1\over2}||1]\cr\hline 
\bl e^*&[1||0]\cr\hline 
\end{array}~\hbox{ and }~
\begin{array}{|c|c|}\hline
&\U(2)\cr\hline\hline
\bl l&[-{1\over2}||1]\cr\hline 
\bl e&[-1|0]\cr\hline\end{array} \cr
\cr
\hbox{fundamental }y=\pm {r\over2},~~r=1,2
\end{eq}The fermion isosinglet fields of the standard model
realize the $2\cdot 3$  fundamental 
$\U(3)$-representations with colour triplet, antitriplet and singlet 
 \begin{eq}{c}
\rep\U(3)=\{[y||2C_1,2C_2]\mid 
y=z+{2(C_1-C_2)\over3},~ z\in\Z,~2C_{1,2}\in\N\}\cr\cr
\begin{array}{|c|c|}\hline
&\U(3)\cr\hline\hline
\bl d&[-{1\over3}||1,0]\cr\hline 
\bl u^*&[-{2\over3}||0,1]\cr\hline 
\bl e&[-1||0,0]\cr\hline\end{array} ~\hbox{ and }~
\begin{array}{|c|c|}\hline
&\U(3)\cr\hline\hline
\bl d^*&[{1\over3}||0,1]\cr\hline 
\bl u&[{2\over3}||1,0]\cr\hline
\bl e^*&[1||0,0]\cr\hline 
\end{array} \cr
\cr
\hbox{fundamental }y=\pm {r\over3},~~r=1,2,3
\end{eq}

Fundamentality for representations 
of the group $\U(2\x3)$ coincides with
standard model fundamentality:
The totally antisymmetric powers of the two defining cyclic fundamental
$\U(2\x 3)$-representations
\begin{eq}{c}
[y||2T;2C_1,2C_2]=\left\{\begin{array}{l}
[+{1\over2\cdot 3}||1;1,0]\cr
[-{1\over2\cdot 3}||1;0,1]\cr\end{array}\right.\cr
\end{eq}give rise to the 
$2\cdot 5$ fundamental $\U(2\x3)$-representations,
which are realized by the
$2\cdot 5$ fundamental fermion fields in the standard model (section 1)
\begin{eq}{c}\hskip-4mm
\rep\U(2\x3)=\{[y||2T;2C_1,2C_2]\mid y=z+T+{2(C_1-C_2)\over3},~
z\in\Z,~2T,2C_{1,2}\in\N\} \cr\cr
\begin{array}{|c|c|}\hline
&\U(2\x3)\cr\hline\hline
\bl q&[{1\over6}||1;1,0]\cr\hline 
\bl d^*&[{1\over3}||0;0,1]\cr\hline 
\bl l^*&[{1\over2}||1;0,0]\cr\hline 
\bl u&[{2\over3}||0;1,0]\cr\hline 
\bl e^*&[1||0;0,0]\cr\hline 
\end{array} 
~~~\hbox{ and }~~~
\begin{array}{|c|c|}\hline
&\U(2\x3)\cr\hline\hline
\bl q^*&[-{1\over6}||1;0,1]\cr\hline 
\bl d&[-{1\over3}||0;1,0]\cr\hline 
\bl l&[-{1\over2}||1;0,0]\cr\hline 
\bl u^*&[-{2\over3}||0;0,1]\cr\hline 
\bl e&[-1||0;0,0]\cr\hline 
\end{array}\cr
\cr
\hbox{fundamental }y=\pm {r\over6},~~r=1,2,3,4,6
\end{eq}The 5th power with hypercharge number ${5\over6}$ is 
not called  fundamental 
since it is a positive linear combination of the 2nd and 3rd
one
\begin{eq}{l}
[{5\over6}||1;0,1]=[{1\over3}||0;0,1]+[{1\over2}||1;0,0]
\end{eq}

The irreducible $\U(2\x3)$-representations can be embedded
by hypercharge renormalization $6y$ as a true submonoid in the representations
of
the direct product group 
\begin{eq}{l}
\rep\U(2\x3)\inmap \rep[\U(1)\x\SU(2)\x\SU(3)]
\end{eq}which should be seen in parallel to  the submonoids for
grand unified schemes, e.g.
$\rep\SU(5)\subnoteq \rep[\U(1)\x\SU(2)\x\SU(3)]$.
The hypercharge number, up to an integer
determined by the nonabelian properties, leads to the fractionality condition
for the standard model fields as given in section 1. 

\section{Integer Charges and Fractional Hypercharges}

The irreducible $\U(1)$-representations as $\U(1)$-endomorphisms	
\begin{eq}{l}	
\U(1)\map\U(1),~~\exp i\al\mape \exp iy\al\cr
[y]\in\rep\U(1)\then y=z\in\Z\cr
\end{eq}have to use 
  integer winding numbers (charge numbers)
  $y\in\Z$ because of  $\exp i(\al+ z2\pi)=\exp i\al$. Since the 
  $\U(\bl1_n)$ subgroup
in $\U(n)$ is synchronized with $\SU(n)$, fractional hypercharge numbers are
possible for  $\U(n)$-representations 
\begin{eq}{l}  
[y||2J_1,\dots 2J_{r-1}]\in\rep\U(n)
\then y=z+{1\over n}{\SUM_{r=1}^{n-1}} r2J_r\in{1\over n}\Z\cr
\exp 2\pi i y\in\I(n)\cong\U(\bl1_n)\cap\SU(n)\cr
\end{eq}

If, somewhere, the group  $\U(1)$ arises as a direct factor, e.g. in
a direct product standard model group $\U(1)\x\SU(2)\x\SU(3)$
or in an asymptotic particle symmetry Cartan subgroup of $\U(2\x 3)$
(section 6), its 
winding numbers have to be integer. 
For a direct product standard model group the hypercharge numbers of
section 1 have to be renormalized to integers $6y$.
$\U(n)$-representations with trivial $n$-ality, 
like 
the fundamental right lepton field $\bl e$, have integer $\U(1)$-winding
numbers from the beginning. 
To attribute integer charge numbers to fields with fractional 
hypercharge numbers, like 
to the fundamental left lepton field $\bl l$ or to the quark fields 
$\bl q,\bl u,\bl d$ (section 1), these field have to be modified, as done
in the Higgs and confinement induced rearrangement of the standard model 
fields to
particle related fields (section 6).  

\section{Normalization of $\U(n)$-Lie Algebras}

The central connection of the internal groups relates to each other also the
normalization of their invariant metrics which arise in the gauge field
coupling constants. 

All complex representations of a real Lie group come with 
an invariant conjugation, e.g. the $\SU(n)$ and 
$\U(n)$-representations spaces
with a defining definite scalar product 
\begin{eq}{l}
V_n\x V_n\map\C,~~\sprod{g\m v}{ g\m w}=\sprod v w\hbox{ for all }g\in \U(n) \cr  
\end{eq}e.g. with  orthonormal
bases $\{e^A\}_{A=1}^n$ and $\{ e^*_A\}_{A=1}^n$ for $V_n$ and 
its dual space $V_n^*$
resp. 
\begin{eq}{l}
\sprod{e^A}{e^B}=\de^{AB},~~
\sprod{ e^*_A}{ e^*_B}=\de_{AB}
\end{eq}

A scalar product for a representation
defines by its powers 
a scalar product for the product representations. 
The  scalar product of the cyclic $\U(n)$-representations 
$[{1\over n}||1,0,\dots,0]$ on $V_n$ and 
$[-{1\over n}||0,\dots,0,1]$ on $V_n^*$ 
induces a scalar product to the 
product space
\begin{eq}{l}
(V_n\ox V_n^*) \x 
(V_n\ox V_n^*)\map\C,~~
\sprod{e^A\ox e^*_B}
{e^C\ox e^*_D}=\de^{AC}\de_{BD}
\end{eq}$V_n\ox V_n^*\cong\C^{n^2}$ is the
$n^2$-dimensional 
defining space  for the group $\U(n)$ and its Lie algebra $\log\U(n)$.
A reordering  gives the metric for
both the abelian 
Lie subalgebra $\log\U(\bl 1_n)$ 
with basis $\{i\bl 1_n\}$ and the 
nonabelian one  $\log\SU(n)$ with 
a basis $\{i\rvec \tau(n)\}$ of generalized
Pauli matrices (Pauli matrices proper  $\{i\rvec\tau\}$ 
for $n=2$, Gell-Mann matrices $\{i\rvec \la\}$ 
for $n=3$ etc.) 
\begin{eq}{l}
V_n\ox V_n^*\supnoteq\log\U(n)=\log\U(\bl 1_n)\pl\log\SU(n)\cr
\de^{AC}\de_{BD}=\left\{
\begin{array}{ll}
\de^A_B\de^C_D&\hbox{for }\U(1)\cr
{1\over2}\de^A_B\de^C_D+{1\over 6}\rvec\tau^A_B\rvec \tau^C_D 
&\hbox{for }\U(2)\cr
{1\over3}\de^A_B\de^C_D+{1\over 12}\rvec \la^A_B\rvec\la^C_D 
&\hbox{for }\U(3)\cr
{1\over n}\de^A_B\de^C_D+{1\over n(n+1)}\rvec\tau(n)^A_B\rvec\tau(n)^C_D 
&\hbox{for }\U(n),~n\ge2\cr\end{array}\right.
\end{eq}This
involves as relative normalization of both 
 Lie subalgebras
\begin{eq}{l}
n\ge2:~~{\norm{\log \SU(n)}^2\over\norm{\log\U(\bl 1_n)}^2}
={\sprod{\tau(n)}{\tau(n)}\over\sprod {\bl 1_n}{\bl 1_n}}=
{n(n+1)\over n}=n+1
\end{eq}The absolute normalization is not determined.

The  coupling constants in the gauge field-current couplings
of the standard model
 \begin{eq}{l}  
g_1\bl A \bl J(1) +g_2\bl B \bl J(2)+ g_3\bl G \bl J(3)\cr
\bl J(1)={1\over6}[\bl q^*\bl 1_6\bl q-2\bl d^*\bl 1_3\bl d 
-3\bl l^*\bl 1_2\bl l +4\bl u^*\bl 1_3\bl u -6\bl e^*\bl e]\cr 
\bl J(2)={1\over2}[\bl q^*\rvec\tau\ox \bl1_3\bl q
+\bl l^*\rvec\tau\bl l]\cr
\bl J(3)={1\over2}[\bl q^*\bl1_2\ox\rvec\la\bl q+\bl d^*\rvec\la\bl d 
 +\bl u^*\rvec\la\bl u] 
\end{eq}have as relative normalizations  in a $\U(2\x 3)$-formulation 
\begin{eq}{l}
\sprod{\bl1}{\bl1}:\sprod\tau\tau:\sprod\la\la=g_1^2:g_2^2:g_3^3=1:3:4
\end{eq}leading to the tree value for the Weinberg angle
\begin{eq}{l}
\tan^2\phi={g_1^2\over g_2^2}={1\over3},~~\sin^2\phi={1\over4}
\end{eq}

\section{Definition of the Electromagnetic Group}

In the following, particles are defined with Wigner \cite{WIG} 
as irreducible unitary
representations of the Poincar\'e group, which includes
 the representations of the
translations. With such a  definition, the 
 quark fields $\bl q,\bl u,\bl d$, parametrizing strong interactions and
hadrons, 
give not rise to asymptotic quark particles if they are confined and, 
therewith, if they cannot develop  translation degrees of freedom.

The transition from the standard model fields to the standard model particles
requires a special basis (ground state) to measure the eigenvalues for the
particle properties.
It is the definition of a symmetry to distinguish no special basis, e.g. it
does not make sense to distinguish an upper and lower component in the left
lepton field $\bl l$ with $\U(2)$-symmetry. However, 
 a basis with upper and lower
component ${\scriptsize\pmatrix{\nu\cr \ro e_L\cr}}$ is necessary to define the
neutrino and electron particle for a lepton field.

The discrete $\I(2)\x\I(3)$-correlation of the hypercharge group  with 
the isospin and colour group  influences strongly
the symmetry breakdown structure which establishes eigenvector bases
for the  particles.
The  transition from the fields with $\U(2\x3)$-properties to  particles,
e.g. 
$\bl l\mape{\scriptsize\pmatrix{\nu\cr \ro e_L\cr}}$,  
has to take into account 
a maximal subgroup of
diagonalizable operators as subgroup of $\U(2\x3)$. 
For a dynamics with a degenerate ground state as in the standard model,
the asymptotic space has to take care, in addition,  of the 
ground state frozen symmetries, parametrized in the standard model
by the ground state properties of a Higgs field $\bl H$ 
as a fundamental $\U(2\x3)$-representation  $[-{1\over2}||1;0,0]$,
trivial for colour $\SU(3)$ and  cyclic
$[-{1\over2}||1]$ for $\U(2)$.
From the  internal real 12-dimensional Lie group 
operators $\U(2\x 3)$ for interactions, there remains
only an abelian electromagnetic $\U(1)$-symmetry for particles. 
In contrast to the fields in the basic dynamics, all particles 
have trivial
isospin symmetry, therefore, the electron and its neutrino can have 
 different mass and different electromagnetic
charge,  and  - if confinement is true - all particles behave trivially
with respect to colour transformations. Since the hypercharge $\U(1)$-subgroup 
in $\U(2\x 3)$ is nontrivially 
synchronized 
both  with  isospin $\SU(2)$ using $\I(2)$  and  with colour $\SU(3)$ using $\I(3)$, 
the definition of an abelian electromagnetic symmetry group
 $\U(1)\subnoteq\U(2\x3)$
with nontrivial hypercharge contributions
has to sever the relation to  both nonabelian factors.

The distinction of an electromagnetic charge
$\U(1)$ in the electroweak hyperisospin $\U(2)$, 
orthogonal to the Higgs ground state expectation value defined direction
\begin{eq}{l}
  \angle{\bl H}={\scriptsize\pmatrix{0\cr m_0\cr}}\ne0,~~
   \exp i\al_+{\bl 1_2+\tau^3\over2}\in\U(1)_+\subnoteq\U(2)
\end{eq}determines uniquely
an electromagnetic group $\U(1)$ only for colour singlets $[z+T||2T;0,0]$
\begin{eq}{l}
\U(2)\cong {\U(1)\x\SU(2)\over\I(2)}~\stackrel{\angle{\bl H}\ne0}
{\longrightarrow}~
\U(1)_+\cong\U(1)
\end{eq}$\angle{\bl H}\ne0$  does not cut the $\I(3)$-relation between 
hypercharge  $\U(1)$ 
and colour $\SU(3)$ in $\U(2\x3)$. The  electroweak breakdown by the Higgs field 
with the definition of the electromagnetic charge $\U(1)$ for particles 
makes sense only together with a colour
confinement
\begin{eq}{l}
 \U(2\x3)\begin{array}{c}
 {\scriptstyle\angle{\bl H}\ne0}\cr
 \longrightarrow\cr
 \hbox{\scriptsize confinement}\end{array} \U(1)
 \end{eq}A colour confinement
 serves simultaneously two things: It trivializes $\SU(3)$ and it 
 allows the definition of an electromagnetic $\U(1)$-group.  In a $\U(3)$-symmetric dynamics, 
quark triplet and antitriplet fields  have both nontrivial hypercharge
and colour properties, 
to attribute to them a unique  electromagnetic charge $\U(1)$ 
does not make sense.

A Cartan Lie subalgebra defines eigenvectors with 
eigenvalues for the `infinitesimal' 
Lie algebra action.
The exponential of a Cartan Lie algebra
in the Lie group under consideration  gives an abelian  subgroup.
 A product of two abelian groups with nontrivial
common subgroup allows no independent measurement of both factors.
Therefore, a Cartan subgroup of a Lie group will be defined as 
a maximal subgroup of a Cartan Lie algebra exponential which can be written
as a direct product of 1-dimensional
Lie groups.
This definition
 of a    Cartan subgroup as a 
 maximal diagonalizable direct product group
 is convenient for the unitary groups, there exist 
 other definitions \cite{WAR}. 
The dimension of a Cartan subgroup need not  coincide with 
the rank of a Lie algebra and, therewith,
with  the dimension of a Cartan Lie algebra.

The exponentials of Cartan Lie algebras for $\SU(n)$  and $\U(n)$
are real $(n-1)$ and $n$-dimensional tori
$\U(1)^{n-1}$ and $\U(1)^n$ resp.
With a diagonal $\SU(2)$- Cartan subgroup
\begin{eq}{l}
\SU(2)\supnoteq\U(1)_0\ni \exp {i\over2}\be_3\tau^3
\end{eq}one obtains diagonal $\SU(n)$-Cartan subgroups as exponentials
\begin{eq}{rl}
\SU(n)\supnoteq&\exp {i\over2}{\SUM_{r=2}^n}\be_r\tau(n,r)=
\begin{array}{c}
{\scriptstyle n}\cr
\X\cr
{\scriptstyle r=2}\end{array}
\U(1)_{0r},~~\U(1)_{0r}\cong\U(1)
\cr
\hbox{for }\SU(2):&
\tau(2,2)=\tau^3={\scriptsize\pmatrix{1&0\cr0&-1\cr}}\cr
\hbox{for }\SU(3):&
\tau(3,2)={\scriptsize\pmatrix{1&0&0\cr0&-1&0\cr 0&0&0\cr}},~
\tau(3,3)={\scriptsize\pmatrix{1&0&0\cr0&0&0\cr 0&0&-1\cr}}
\end{eq}

Cartan subgroups of $\U(n)=\U(\bl1_n)\o\SU(n)$
 have to take care of the central $\I(n)$-correlation,
e.g. the $\U(1)$-isomorphic subgroups
$\exp{i\over2}\al\bl1_2\in\U(\bl1_2)$ and
$\exp{i\over2}\be_3\tau^3\in\U(1)_0$ 
form a product subgroup $\U(\bl1_2)\o\U(1)_0\subnoteq\U(2)$, but with 
the nontrivial intersection $\exp i\pi\bl1_2=\exp i\pi\tau^3=-\bl1_2$,
no  direct product $\U(1)\x\U(1)$  in $\U(2)$.
A  $\U(n)$-Cartan subgroup 
with appropriate Lie parameters is given by
\begin{eq}{l}
\U(n)\supnoteq
\exp i{\SUM_{r=1}^n}\al_r \bl1(n,r)
=\begin{array}{c}
{\scriptstyle n}\cr
\X\cr
{\scriptstyle r=1}\end{array}
\U(1)_r,~~\U(1)_r\cong\U(1)
\end{eq}using a  maximal system of $n$
orthogonal projectors for the Lie algebra $\log\U(n)$
\begin{eq}{l}
\{\bl1(n,r)\mid r=1,\dots n\}\hbox{ with } 
\left\{\begin{array}{l}
\bl1(n,r)\o \bl1(n,s)=\de_{rs}\bl1(n,r)\cr
{\SUM_{r=1}^n}\bl1(n,r)=\bl1_n\end{array}\right.\cr
\end{eq}e.g. with matrices $\bl1(n,r)$ having only one nontrivial diagonal entry  
\begin{eq}{rl}
\hbox{for }\U(2):&
\bl1(2,1)={\bl 1_2+\tau^3\over2}={\scriptsize\pmatrix{1&0\cr0&0\cr}},~~
\bl1(2,2)={\bl 1_2-\tau^3\over2}={\scriptsize\pmatrix{0&0\cr0&1\cr}}\cr
\hbox{for }\U(3):&
\bl1(3,1)={\scriptsize\pmatrix{1&0&0\cr0&0&0\cr 0&0&0\cr}},~
\bl1(3,2)={\scriptsize\pmatrix{0&0&0\cr0&1&0\cr 0&0&0\cr}},~~
\bl1(3,3)={\scriptsize\pmatrix{0&0&0\cr0&0&0\cr 0&0&1\cr}}
\end{eq}Both for $\SU(n)$ and $\U(n)$ the rank $(n-1)$ and $n$ resp.
gives also the Cartan subgroup dimension.

This is different for the rank 4 standard group
$\U(2\x3)=\U(\bl1_6)\o[\SU(2)\ox\bl1_3\x\bl1_2\ox\SU(3)]$ where
the Cartan subgroups with direct $\U(1)$-factors 
have  to be looked for in the real 4-dimensional 
exponential of a Cartan Lie algebra
\begin{eq}{l}
\exp i[{\al\over6}\bl 1_6+{\be_3\over2}\tau^3\ox\bl1_3
+\bl 1_2\ox {\ga_3\la^3+\ga_8\la^8\over2}]\in\U(1)^4
\end{eq}In a Cartan subgroup, hypercharge $\exp i{\al\over6}\bl 1_6$ has to be
correlated either with isospin  or with  colour. 
Correspondingly, there arises 
two systems - one  with  two and  one with three 
orthogonal projectors for $\U(2)$ and $\U(3)$ resp. 
\begin{eq}{rl}
\hbox{hyperisospin:}&
\left\{\begin{array}{rl}
\hbox{projectors: }&\{\bl 1(2,r)\ox\bl 1_3\mid r=1,2\}\cr
\hbox{Cartan subgroup: }&{\scriptsize\pmatrix{
\exp i\al_+&0\cr
0& \exp i\al_-\cr}}\ox\bl 1_3\cr
&\in\U(1)_+\x\U(1)_-\cr\end{array}\right.\cr
\hbox{hypercharge-colour:}&
\left\{\begin{array}{rl}
\hbox{projectors: }&\{\bl 1_2\ox\bl 1(3,r)\mid r=1,2,3\}\cr
\hbox{Cartan subgroup: }&\bl 1_2\ox {\scriptsize\pmatrix{
\exp i\al_1&0&0\cr
0& \exp i\al_2&0\cr
0&0& \exp i\al_3\cr
}}\cr&\in\U(1)_1\x\U(1)_2\x\U(1)_0\cr\end{array}\right.\cr
\end{eq}The two Cartan subgroups have dimension two and three resp.

For the characterization of the
asymptotic particle states for the fields in the
 $\U(2\x3)$-symmetric standard model 
as eigenvectors with eigenvalues of a maximal diagonal
subgroup, 
one has to decide either 
for $\U(1)_1\x\U(1)_2\x\U(1)_3\subnoteq\U(3)$
leading to isospin trivial particles with possibly nontrivial
hypercharge-colour properties or
for $\U(1)_+\x\U(1)_-\subnoteq\U(2)$
leading to  colour singlet particles  with possibly nontrivial
hyperisospin properties as done 
in the standard model. 
By the Higgs field colour singlet property,
the Higgs mechanism
decides  for the Cartan subgroup in hyperisospin $\U(2)$, 
therein it establishes a basis with a remaining  
$\U(1)_+\x\U(1)_-$
and  trivializes, in addition,  one direct factor $\U(1)_-$ which 
leaves nontrivially only  the electromagnetic $\U(1)_+$-symmetry
\begin{eq}{rcl}
\U(2\x3)&\stackrel{\rm confinement}{\map}&{\AND^3}\U(2\x3)\cong 
\U(2)\cr
\U(2)&\stackrel{\angle{\bl H}\ne0}{\map}&
{\U(\bl1_2)\o\U(1)_0\over\U(1)_-}\cong
{\U(1)_+\x\U(1)_-\over\U(1)_-}\cong\U(1)	
\end{eq}
	
\section{The  External-Internal Correlation}

In addition to the internal correlation of hypercharge with
isospin-colour, the standard  dynamics
shows in the Yukawa interaction 
\begin{eq}{l}
(\mu_e\bl e^*\bl l+\mu_u\bl q^*\bl u+\mu_d\bl d^*\bl q)\bl H
+\hbox{h.c.}~~\hbox{ with Yukawa couplings }\mu_{e,u,d}\in\R
\end{eq}an  internal-external correlation
for chirality (left and right handedness)
and hyperisospin. The 
Higgs field, which connects left and right handed fields,
is a hyperisospin doublet.

Also the  Lorentz transformations, defined
by the covering group $\SL(\C^2 )$ with $\centr\SL(\C^2 )\cong\I(2)$ of
the orthochronous group $\SO^+(1,3)$, and being a subgroup of 
the unimodular group $\la\in{\bf UL}(2)\subnoteq \GL(\C^2)$
with $|\det \la|=1$, have  an 
$\I(2)$-correlation with the external phase group $\U(\bl1_2)\subnoteq\GL(\C^2)$
\begin{eq}{l}
{\bf UL}(2)=\U(\bl1_2)\o\SL(\C^2 ),~\left\{
\begin{array}{l}
\U(\bl1_2)\cap\SL(\C^2 )\cong\I(2)\cr
{\bf
UL}(2)/\U(\bl1_2)\cong\SL(\C^2 )/\I(2)\cong\SO^+(1,3)\end{array}\right.\cr
{\bf UL}(2)\cong{\U(1)\x\SL(\C^2 )\over\I(2)}
\end{eq}The finite dimensional 
irreducible  ${\bf UL}(2)$-representations
\begin{eq}{c}
\rep{\bf UL}(2)=\{[c||2J_L|2J_R]\mid c=z+J_L-J_R,~ z\in\Z,~ 2J_{L,R}\in\N\}\cr
[c||2J_L|2J_R]^*=[-c||2J_R|2J_L]
\end{eq}are characterized by left and right spin winding numbers $2J_{L,R}$
and, in addition, 
a chirality number $c$,
characterizing the representation of $\U(\bl1_2)\subnoteq
\bl{UL}(2)$. 
Integer spin fields have integer
chirality numbers $c$, 
halfinteger spin comes with
halfinteger chirality.

From the $\U(2\x3)$ and ${\bf UL}(2)$-invariant Yukawa interaction, 
the ${\bf UL}(2)$-properties 
can be attributed to the fundamental fields as follows
\begin{eq}{c}
\begin{array}{|c|c|}\hline
&{\bf UL}(2)\cr\hline\hline
\bl l&[+{1\over2}+z_l||1|0]\cr\hline 
\bl e&[-{1\over2}+z_l||0|1]\cr\hline 
\bl q&[+{1\over2}+z_q||1|0]\cr\hline 
\bl d&[-{1\over2}+z_q||0|1]\cr\hline 
\bl u&[+{3\over2}+z_q||0|1]\cr\hline 
\bl H&[-1||0|0]\cr\hline 
{\bf A,B,G}&[0||1|1]\cr\hline 
\end{array}~~ \hbox{ with } z_l,z_q\in\Z 
\end{eq}With chirality number $c=-1$ fixed for the Higgs field,
the chirality numbers
of quark and lepton fields are determined up to integers.

A
 first suggestion for a central
external-internal correlation 
could be an isospin-spin correlation as expressed by
the quotient group ${\SL(\C^2 )\x\SU(2)\over\I(2)}$, leading
to the correlation half integer spin with half integer isospin and integer
spin with integer isospin as seen in the lepton and quark isodoublet
fields $\bl l$ and $\bl q$ or the gauge fields $\bl A$, $\bl B$ and $\bl G$
resp.
However, the integer isospin, half integer spin fermion  fields
$\bl e$, $\bl u$ and $\bl d$ and the Lorentz scalar isodoublet Higgs field     
$\bl H$ 
contradict such a suggestion. The external-internal correlation is different.

The  internal hypercharge $\U(1)\subnoteq\U(2\x3)$  
 and the external chirality $\U(1)\subnoteq\UL(2)$, have to use
 the same phase - both $\U(1)$ coincide in the Higgs field
 with the vanishing combination $c-2y$. This gives
with the choice
\begin{eq}{l}
z_l=-2,~~z_q=0
\end{eq}as $\U(1)$-properties 
of the standard model fields
\begin{eq}{c}
\begin{array}{|c|c|c|c|}\hline
&\U(1)_{\rm ext}&\U(1)_{\rm int}&\U(1)_{\rm ferm}\cr
&c&y&c-2y\cr\hline\hline
\bl e&-{5\over2}&-1&-{1\over2}\cr\hline 
\bl l&-{3\over2}&-{1\over2}&-{1\over2}\cr\hline 
\bl d&-{1\over2}&-{1\over3}&+{1\over6}\cr\hline 
\bl q&+{1\over2}&{1\over6}&+{1\over6}\cr\hline 
\bl u&+{3\over2}&{2\over3}&+{1\over6}\cr\hline 
\bl H&-1&-{1\over2}&0\cr\hline 
{\bf A,B,G}&0&0&0\cr\hline 
\end{array} 
\end{eq}The correlation of external and
internal $\U(1)$ defines a fermion number group $\U(1)$ 
\begin{eq}{l}
\left.\begin{array}{rl}
\U(1)_{\rm ext}&\subnoteq{\bf UL}(2)\cr
\U(1)_{\rm int}&\subnoteq\U(2\x3)\end{array}\right\}
\U(1)_{\rm ferm}\cong{\U(1)_{\rm ext}\o\U(1)_{\rm int}\over\U(1)}\cr
f=c-2y=\left\{\begin{array}{rl}
-{1\over2}&\hbox{for lepton fields }\bl e,~\bl l\cr
+{1\over6}&\hbox{for quark fields }\bl d,~\bl q,~\bl u\end{array}\right.
\end{eq}For the left handed fields $\bl l,\bl q$, the fermion number
$\U(1)$ coincides with the hypercharge $\U(1)$, i.e. $c=3f=3y$.

Taking into account the identification of the external and
internal phase group, the symmetry group,
faithfully represented in the
standard model,  	
shows three central correlations - the discrete internal one by $\I(2)\x\I(3)$,
the discrete  external one by $\I(2)$ and, finally, the continuous 
external-internal one
by $\U(1)$
\begin{eq}{l}
{\U(1)_{\rm ext}\x\U(1)_{\rm int}\x\SL(\C^2 )\x\SU(2)\x\SU(3)
\over\U(1)\x\I(2)\x\I(2)\x\I(3)}
\cong{{\bf UL}(2)\x\U(2\x3)\over\U(1)}
\end{eq}It's as complicated!

The representations of the external-internal group 
by the standard model fields are summarized  as follows
\begin{eq}{c}
\begin{array}{|c||c|c|}\hline
\hbox{field}&
 &{\bf UL}(2)\o\U(2\x3)\cr
 &&[c||2J_L|2J_R]\o[y||2T;2C_1,2C_2]\cr
  \hline\hline
\hbox{right lepton}&
 \bl e&[-{5\over2}||0|1]\o[-1||0;0,0]\cr\hline
\hbox{left lepton}&
\bl l&[-{3\over2}||1|0]\o[-{1\over2}||1;0,0]\cr\hline
\hbox{right down quark}&
\bl d&[-{1\over2}||0|1]\o[-{1\over3}||0;1,0]\cr\hline
\hbox{left quark}&
\bl q&[+{1\over2}||1|0]\o[+{1\over6}||1;1,0]\cr\hline
\hbox{right up quark}&
\bl u&[+{3\over2}||0|1]\o[+{2\over3}||0;1,0]\cr\hline
\hbox{Higgs}&
\bl H&[-1||0|0]\o[-{1\over2}||1;0,0]\cr\hline
\hbox{hypercharge gauge}&
\bl A&[0||1|1]\o[0||0;0,0]\cr\hline
\hbox{isospin gauge}&
\bl B&[0||1|1]\o[0||2;0,0]\cr\hline
\hbox{colour gauge}&
\bl G&[0||1|1]\o[0||0;1,1]\cr\hline
\end{array}
\end{eq}

\section{The Complication of the\\ Standard Model Group}

From an esthetical standpoint, debatable of course, the  
external-internal group
$\U(1)\x\SL(\C^2 )\x\U(1)\x \SU(2)\x\SU(3)$ with five direct factors 
(chirality, Lorentz symmetry, hypercharge, isospin, colour)
may look rather unnatural and complicated, even more
its quotient group above with the four central correlations.
Grand unified theories hope for a 
correlation of the direct factors via `nondiagonal' supplements in
larger simple groups,
like $\SU(5)$ or $\SO(10)$ 
for the internal symmetry, where, however, if no additional benefits arise, 
  the symmetry breakdown mechanism for
the asymptotic particles with  a leftover electromagnetic $\U(1)$ 
destroys the simplification hoped for.

The fundamental standard model fields are rather close to their
associated asymptotic particles. With the important exceptions
of the hadrons, requiring  colour
confinement, and the gauge fields related particles,
 requiring gauge invariance, there is even a bijective correpondence 
between interaction parametrizing fields
and particle fields. This is similar to the quantum mechanical harmonic
oscillator where the position-momentum operators $(\ro x,\ro p)$ are
linearily and bijectively related to the 
energy eigenstates defining creation-annihilation pair 
$(\ro u,\ro u^*)$ with $\ro u={\ro x+i\ro p\over\sqrt2}$. 
For a quantum mechanical dynamics with a Hamiltonian,
not written with energy eigenvectors, 
like for the nonrelativistic hydrogen atom,
the connection between the dynamics building operators $(\ro x,\ro p)$
and the energy eigenstates generating operators are nonlinear and may be rather  
complicated. Could it be that the  standard model group complication arises
by its close relationship to the asymptotic particles?
Is the complicated standard model symmetry a consequence of a
linearization with  many 
energy-momentum and electromagnetic eigenfields 
to approximate a nonlinear dynamics, which can be formulated 
with a few quantum fields implementing a smaller symmetry?
Does the complicated internal group of the standard model express 
 not a subgroup of a larger symmetry group, but 
a representation spectrum of a smaller group?
I will close this paper with some speculations 
for such a small unification, in some sense orthogonal to those aiming at
grand unification schemes.

As seen in the hypercharge-isospin-colour correlation,
the full unitary group $\U(n)$ has a  root structure with respect to $\U(1)$
and the totally antisymmetric product
in analogy to the cyclotomic group $\I(n)$ as its centrum 
having a root structure with respect  to the trivial group $\{1\}$ and the number product
\begin{eq}{l}
{\AND^n}\U(n)\cong \U(1),~~(\I(n))^n\cong\{1\}
\end{eq}Obviously, such a structure is used for the 
baryonic hadronization ${\AND^3}\ro q\cong\ro N$ with
the  quarks  as `cubic roots' of the nucleon.

For relatively prime $n,m\in\N$, one has
the root structure for $\U(n\x m)$ as defined as subgroup of $\U(nm)$
\begin{eq}{l}
{\AND^m}\U(n\x m)\cong \U(n),~~\(\I(n)\x\I(m)\)^m\cong \I(n)
 \end{eq}Such a property holds for the 
 centrally connected  standard model group $\U(2\x 3)$,
 not, however,  for 
 the direct product group $\U(1)\x\SU(2)\x\SU(3)$ since ${\AND^m}\U(1)\cong\{1\}$ for
 $m\ge2$.
 
If there is a  
 dynamics underlying the standard model, 
 given by an invariant of the full unitary group $\U(2)$ and built
with the defining  representation, carried by fermion fields  
\begin{eq}{rll}
 \psi,\psi^*&\cong
 [\mp{1\over2}||1]&\in\rep\U(2)\cr
 \psi\and\psi,\psi^*\and\psi^*&\cong
 [\mp1||0]&\in\rep\U(1)\cr
 \psi\and\psi\ox\psi^*\and\psi^*&\cong
 [0||0]&\in\rep\{1\}\cr
\end{eq}it may give rise to energy-momenta eigenstates in the 
basic field products,
e.g. in
\begin{eq}{l} 
 \psi\ox  \psi^*\and  \psi^*  \cong[{1\over2}||1]\in\rep\U(2)
\end{eq}The occurence of energy-momentum eigenstates, created by   
the original cyclic $\psi$ and
 its  product $ \psi\ox  \psi^*\and  \psi^*$,
 may be parametrized by introducing two asymptotically oriented fields,
related to leptons and quarks, the quarks  carrying
an additional $\SU(3)$ as a custodian  symmetry which  expresses their
cubic root origin and distinguishes them from
the leptons
\begin{eq}{rcll} 
 \psi\sim & \bl l&\cong[-{1\over2}|1;0,0]&\in\rep\U(2)\cr 
  \psi\ox  \psi^*\and  \psi^*\sim&{\AND^3}\bl  q
&\cong[{1\over2}|1;0,0]&\in\rep\U(2)\cr 
& \bl q
&\cong[{1\over6}|1;1,0]&\in\rep\U(2\x3)\cr 
\end{eq}

With the introduction of a spectrum describing group for
the Fermi fields,
an asymptotic parametrization 
for 
energy-momentum eigenstates also in the bosonic  
adjoint $\U(2)/\U(\bl1_2)$-re\-pre\-sen\-ta\-tions 
might be necessary,
introducing hypercharge, isospin and colour fields 
\begin{eq}{rcll}
 \psi\ox  \psi^*\sim&\bl   A,~ \bl  B
&\cong[0|0;0,0],~[0|2;0,0]&\in\rep\U(2)\cr 
&\bl G&\cong[0|0;1,1]&\in\rep\U(2\x3)\cr 
\end{eq}

Obviously with ${\AND^3}\U(2\x3)\cong\U(2)$, the basic representation 
structure  is embedded,
$\rep\U(2)\inmap\rep\U(2\x3)$.
The common origin of all the particle oriented fields, 
introduced for the linearization of the basic dynamics, 
remains visible in the hypercharge $\U(1)$-relation  
both to the centrum $\I(2)$ of isospin $\SU(2)$ and 
to the centrum $\I(3)$ of 
the colour symmetry $\SU(3)$.

A substantation of such speculations, as sketched
on the group theoretical representation level only, even
without discussing the external-internal correlation,
 requires the solution of a quantum field
theoretical bound state problem which seems to be extremely difficult, 
as known from the attempts to determine the hadronic spectrum from quantum
chromodynamics.

 \end{document}